\documentclass[twoside,leqno,twocolumn]{article}

\usepackage[letterpaper]{geometry}

\usepackage{ltexpprt}

\usepackage{booktabs} 
\usepackage{xcolor,colortbl}
\usepackage{supertabular}
\usepackage{numprint}
\usepackage{afterpage}
\usepackage{float}
\usepackage{stackengine}
\usepackage{caption}
\usepackage[list=true]{subcaption}
\usepackage[inline]{enumitem}
\usepackage{algorithm}
\usepackage{algpseudocode}
\usepackage{tabularx}
\usepackage{physics}
\usepackage{graphicx}
\usepackage{flushend}
\usepackage{multirow}
\usepackage{tabu}
\usepackage{url}
\algtext*{EndWhile}
\algtext*{EndIf}

\begin{document}




\title{Maximum Coverage Optimization of Twitter Search API Queries\\ for Topic Classifiers}
\author{Kasra Safari\thanks{\texttt{kasra.safari@mail.utoronto.ca}, University of Toronto.}
\and Scott Sanner\thanks{\texttt{ssanner@mie.utoronto.ca}, University of Toronto.}}

\date{}

\maketitle







\begin{abstract} \small\baselineskip=9pt 
Recent work has shown that machine learning classifiers trained to detect topical content on Twitter (e.g., content relevant to ``social issues'') can generalize well beyond the training data and provide stable performance over long time horizons. However, in a setting where one does not have access to the Twitter ``firehose'' and is restricted by Twitter search API query limits, applications must use a two stage process and first decide what content to retrieve through the search API before filtering that content with topical classifiers.  Thus, it is critically important to query the Twitter API relative to the intended topical classifier in a way that achieves high precision and recall.  To this end, we propose a sequence of query optimization methods that generalize notions of the maximum set coverage problem to find the subset of query terms within the API limits that retrieve a large fraction of the topically relevant tweets without sacrificing precision.  We evaluate the proposed methods on a large two year Twitter dataset labeled using manually curated hashtags for different topics.  
Our analysis shows that the best proposed method (CAILP)  
(i)~admits an efficient and effective greedy approximation, 
(ii)~significantly reduces the amount of data retrieved from the search API vs. any other method while (iii)~maintaining high recall relative to the firehose and (iv)~achieving comparable or better performance on various precision metrics compared to the original classifier evaluated directly on the firehose.
\end{abstract}

\section{Introduction}

Recent work has shown that machine learning classifiers trained to detect topical content on Twitter can generalize well beyond the training data and provide stable performance over long time horizons~\cite{sanner:topical}.  Such trained topic classifiers can be easily used to filter and rank content relevant to a given topic.  For example, consider the use case where a user wants to follow Twitter content on a specific topic such as ``social issues''.  Every day, or more frequently, the user would like to check for the latest tweets relevant to the topic in real-time and they would like to do so with high recall (i.e., they do not want to miss important content) as well as high precision (i.e., to make best use of their limited time to browse ranked search results).

\begin{figure}
\centering
\includegraphics[width=\columnwidth]{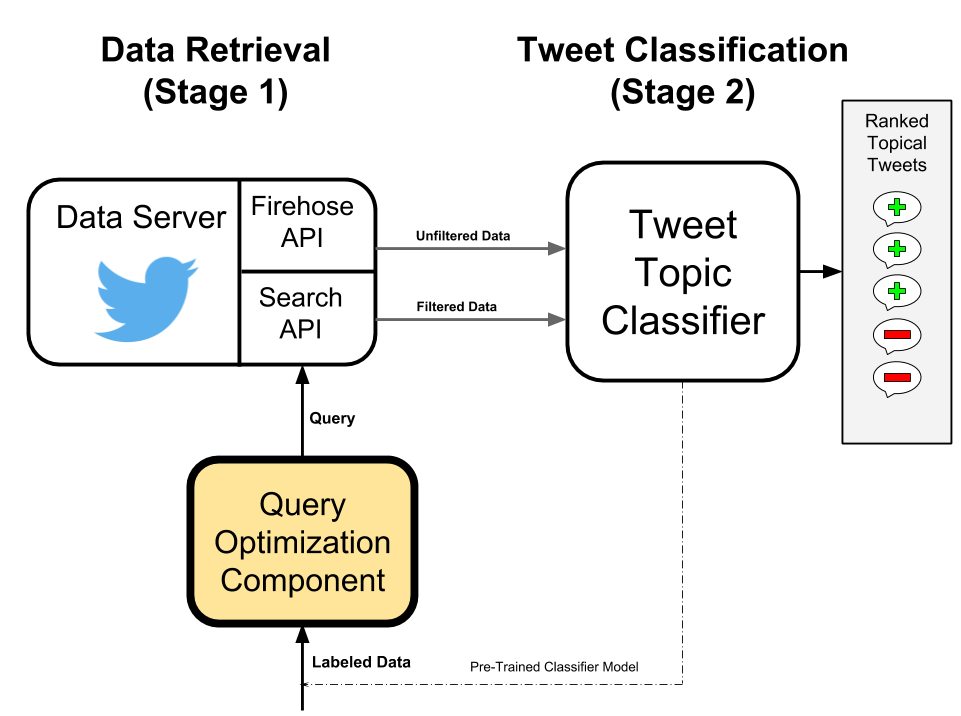}
\caption{This diagram shows possible data flows for real-time tweet classification and ranking given a pre-trained Twitter topic classifier.  If we had access to the Twitter ``firehose'', the data flow would follow the top arrows and we could classify and rank \emph{all} data.  However, if we only had access to the search API, then we can use the classifier-labeled data to select an optimal set of API search queries; the retrieved data filtered by the keywords then flows to the topic classifier and ranker.  Our objective is to optimize selection of API search queries to achieve high recall and precision relative to the gold standard firehose using methods that are intentionally agnostic to the classification approach.}
\label{fig:pipelines}
\end{figure}

Under real-time requirements where one does not have access to the Twitter ``firehose'' (full, unfiltered data streams) and is limited by query restrictions of Twitter's free APIs, one is faced with two main choices: filter a relatively small --- roughly 1\% of the Twitter content (assumed random) \cite{a3d68974958f4ba3a4e9e6b589c02333} --- data  or query for a filtered subset of tweets using specific keywords provided to either of the Twitter search API\footnote{\url{https://developer.twitter.com/en/docs/tweets/search/overview}} or the streaming API\footnote{\url{https://developer.twitter.com/en/docs/tweets/filter-realtime/overview}}, which have limiting constraints on the accepted size of queries (e.g., the search API doesn't accept queries with more than 500 characters).    

In this paper, we propose to use the search API to perform efficient and targeted retrieval of topical content in a two stage framework summarized in Fig.~\ref{fig:pipelines}.  In this framework, a key question is how to formulate first stage queries specific to a given topic in a way that ensures high coverage of relevant content within the search API bandwidth limits.  One approach 
is to use a few manually selected terms or hashtags to construct the search queries \cite{Starbird:2012:RRI:2145204.2145212,Nichols:2012:SSE:2166966.2166999}. However, such predefined queries are subjective in nature and often lead to low recall~\cite{Wang:2013:EHA:2492517.2492624}. 
As a concrete example, we revisit the ``social issues'' topic. Here, we could query for ``\#michaelbrown'' or more generally ``\#blacklivesmatter''; while these are perfectly valid queries, they are inherently high precision but low recall query terms that cover very specific subtopics. In contrast, query terms like ``protesters'' or ``justice'' tend to provide much broader topical coverage that vastly increases recall --- while precision suffers, we'll see this is addressed during ranking in the second stage.  Thus we are left with the key question of how to choose an ``optimal'' subset of query terms 
to achieve high recall at the first stage.

To address optimal query construction in the first stage, we use  query optimization methods that generalize notions of the maximum set coverage problem \cite{KHULLER199939} to find subsets of query keywords that cover the majority of the topically relevant tweets (high recall) while filtering out a large fraction of the irrelevant content (increased precision). We formulate the methods as Integer Linear Programs (ILP) \cite{wolsey2014integer} and show they can be \emph{efficiently} approximated while maintaining performance.

To evaluate our proposed methods, we ran experiments over a large corpus of Twitter data collected over a two year period with seven different trained topic filters.  We use the entire dataset as the firehose and simulate a boolean filtering search API interface to the data for evaluation purposes.  Our results demonstrate the best of our proposed methods (CAILP) achieves competitive recall compared to the firehose, retrieving only a small fraction of the firehose data. Additionally, we show that simple greedy approximations of CAILP solutions perform nearly as well as optimal ILP solver solutions with substantially faster processing times. For the second stage, CAILP shows final ranking performance comparable to classification directly on the firehose, despite retrieving only a small fraction of total firehose tweets at the first stage.


\section{Background}

\subsection{The Two Stage Retrieval and Ranking Framework.}

We use the two stage framework of Fig.~\ref{fig:pipelines}:

\begin{itemize}
\item {\bf 1st Stage -- API query:} Here, we query the API to retrieve the data for ranking at the second stage.  Our main goal for this stage is to maximize the recall --- while we can expect the second stage classifier to weed out the irrelevant data, failure to retrieve enough relevant data in the first stage will negatively affect performance of the second stage.  However, we can't neglect the importance of precision at the first stage because there are rate limits for these APIs\footnote{\url{https://developer.twitter.com/en/docs/basics/rate-limits}} that prevent us from retrieving large amounts of data in feasible times.  Therefore, it is also important to consider a relatively high precision as a secondary objective in the first stage. 

\item {\bf 2nd stage -- Topical classification and ranking:} Using the data retrieved from the API in the first stage, a classifier model (with a score-based output such as a classification probability) is used to rank the tweets for the end-user according to probability of relevance. The main objective at this second stage is to achieve relevant tweets at high ranks.  Common metrics for evaluating user-facing rankings include Average Precision (AP) and Precision@$k$ (P@$k$), both of which we use to empirically evaluate second stage performance.
\end{itemize}
 
\subsection{Maximum K-Coverage Problem.}
Our goal of retrieving topically relevant tweets in the 1st stage given the imposed limitations of the query API has parallels with the well-known maximum $K$-coverage problem from combinatorial optimization, which we now review:

\begin{Definition}[Maximum $K$-Coverage Problem]
Given a set of elements $E$ and a collection of non-empty subsets of $E$ denoted by $S = \{S_1, S_2, \dots, S_m\}$, maximize the number of unique elements of $E$ covered by selecting only $K$ subsets from $S$.
\end{Definition}

Now we show how the task of querying a limited API can be reduced to a maximum $K$-coverage problem. We have a corpus of labeled tweets that can be filtered based on the features present in their content. 
Our task is to select a limited number of these features which are present in the majority of the tweets in the corpus. This can be formulated as a max $K$-coverage problem in which the universal set of elements to be covered is the tweet corpus and the tweet features select candidate subsets of tweets to be covered by the solution.

\vspace{1mm}
\noindent{\bf Notation and Objective.} Before moving on to a  formal definition of our specific proposals, we first define the notation used in the rest of the paper. Since the tweets are labeled, the collection of tweets can be divided into two disjoint sets $P$ and $N$. $P = \{P_i\}$ for $i \in \{ 1 \ldots |P| \}$ is the collection of positively labeled tweets $P_i$ (relevant). Similarly, $N = \{N_i\}$ for $i \in \{ 1 \ldots |N| \}$ is the collection of negatively labeled tweets (irrelevant). 

The set $F = \{F_j\}$ for $j \in \{ 1 \ldots |F| \}$ contains all the candidate features that can be used in the query to filter the tweets. Given a query for an arbitrary feature $F_j$, an ideal API with no limitations on the number of tweets returned will give us all the tweets in the corpus containing $F_j$ (we can represent these tweets with $\mathit{cov}(F_j)$ for every feature).

In summary, our overall objective in the 1st stage is to find an optimal set of $k$ tweet features $F^*$ to be used in a Twitter Search API query that provides high coverage of tweets in $P$ and low coverage of tweets in $N$.  How specifically we define and optimize this objective is covered in the following subsections.

\subsection{ILP Formulations for Query Optimization.}

Integer Linear Programs (ILPs) are commonly used to formalize maximum coverage problems, thus we use ILP notation to formulate variants of our Search API Query optimization objectives.
We begin with a formulation focused on the coverage of the positive labeled data and subsequently augment this ILP with weightings for features as well as additional components for suppressing the irrelevant (negative labeled) data.

In the following ILP formulations, lowercase letters $f, p, n$ are used as binary variables to indicate the presence of features and tweets in the ILP solution. For each feature $F_j \in F$, if $f_j = 1$ the feature is selected to be included in the query. Similarly, $p_i$ and $n_i$ are used for the tweets in $P$ and $N$. $p_i = 1$ indicates that the query covers tweet $P_i$ (and the same for $n_i$ w.r.t.\ $N_i$).

\vspace{1mm}
\noindent \textbf{Coverage-based ILP Formulation (CILP).} For the simplest formulation, we only use the coverage of topical (positive labeled) tweets by each feature:
\begin{equation*}
\begin{aligned}
& \underset{f}{\text{maximize}} 
& & \sum_{i=1}^{|P|}p_i \\
& \text{subject to}  
& & p_i = \bigvee_{\{j:P_i\in cov(F_j)\}}f_j, \; i = 1,\ldots, |P| \\
&&& \sum_{j=1}^{|F|}f_j \le K \\
&&& p_i \in \{0,1\} , \; i = 1,\ldots, |P| \\
&&& f_j \in \{0,1\} , \; j = 1,\ldots, |F| 
\end{aligned}
\end{equation*}

The objective function of this ILP is to maximize the coverage of positive labeled tweets in $P$. This coverage can be calculated by summing up all the binary variables $p_i$ for each tweet $P_i \in P$.  The first constraint ensures that no tweet increases the coverage in the objective function count unless (at least) one of its features is selected in the solution. We use a binary OR for all the features $F_j$ that have the tweet in their coverage to enforce this. The second constraint controls the number of features in the solution to not exceed the predetermined number of $K$ features.

\vspace{1mm}
\noindent \textbf{Weighted ILP Formulation (WILP).} While the CILP formulation offers the most simplicity, a possible issue is that 
a feature may be selected by this model that covers both a large fraction of topical and non-topical tweets, which is not desirable. 
One way to encourage the selection of topically informative features is by weighting them according to their Mutual Information (MI) \cite{Cover:2006:EIT:1146355} with the tweet topic label. 
In this formulation, we use the MI of the features selected in the solution as an additional term in the decision function similar to a regularization term. In this way, selecting features with higher MI is rewarded by an increase in the objective function. The mutual information scores are approximated using the available data during training.

We use the Normalized Mutual Information Score (NMIS)~\cite{estevez2009normalized} for a feature $F_j$ with the topic label, which outputs scores in a range between 0 (no mutual information) to 1 (maximum shared information). The total coverage in the objective function is also normalized by dividing by the total number of tweets ($|P|$) so that both terms in the objective function are similarly scaled:
\begin{equation*}
\begin{aligned}
& \underset{f}{\text{maximize}} 
& & \dfrac{\sum_{i=1}^{|P|}p_i}{|P|} + \lambda\sum_{j=1}^{|F|}f_j \cdot \mathrm{NMIS}(F_j) \\
& \text{subject to}  
& & p_i = \bigvee_{\{j:P_i\in cov(F_j)\}}f_j, \; i = 1,\ldots, |P| \\
&&& \sum_{j=1}^{|F|}f_j \le K \\
&&& p_i \in \{0,1\} , \; i = 1,\ldots, |P| \\
&&& f_j \in \{0,1\} , \; j = 1,\ldots, |F| 
\end{aligned}
\end{equation*}
The importance of the NMIS portion of the objective is controlled using hyperparameter $\lambda \in [0,\infty)$.  Determining the value of $\lambda$ is a hyperparameter tuning task. In our case, we tuned $\lambda$ to maximize the F1-score on held-out validation data to balance recall and precision. 

\vspace{1mm}
\noindent \textbf{Coverage/Anti-coverage Based ILP Formulation (CAILP).} 
An alternative to the WILP for penalizing non-topically predictive features is to directly reward features for coverage of topical tweets as in the CILP, but to also reward anti-coverage of non-topical tweets:
\begin{equation*}
\begin{aligned}
& \underset{f}{\text{maximize}} 
& & \dfrac{\sum_{i=1}^{|P|}p_i}{|P|} - \lambda\dfrac{\sum_{l=1}^{|N|}n_l}{|N|} \\
& \text{subject to}  
& & p_i = \bigvee_{\{j:P_i\in cov(F_j)\}}f_j, \; i = 1,\ldots, |P| \\
&&& n_l = \bigvee_{\{j:N_l\in cov(F_j)\}}f_j, \; l = 1,\ldots, |N|  \\
&&& \sum_{j=1}^{|F|}f_j \le K \\
&&& p_i \in \{0,1\} , \; i = 1,\ldots, |P| \\
&&& n_l \in \{0,1\} , \; l = 1,\ldots, |N| \\
&&& f_j \in \{0,1\} , \; j = 1,\ldots, |F|  
\end{aligned}
\end{equation*}
Here, the negated second term of the objective (and corresponding constraint to define $n_l$) rewards non-coverage of non-topical tweets.  Similar to the WILP, the $\lambda$ hyperparameter balances the two objectives and is tuned on validation data as described for the WILP. 

\subsection{Solving the Integer Linear Programs.}
Maximum $K$-coverage and its variants are typically solved by two standard approaches~\cite{KHULLER199939}:
\begin{enumerate}
\item \textbf{Optimization Solvers.} One option is to use commercially available tools to optimally solve the ILP formulation.  Unfortunately as we will see, their scalability is quite limited. 
\item \textbf{Greedy Algorithms.} Greedy approaches attempt to approximately optimize set coverage problems by iteratively selecting one feature at a time that most improves the objective until $K$ features are selected as in Algorithm~\ref{alg:greedy}.
While not guaranteed to be optimal, greedy methods are highly scalable and known to be capable of producing good solutions for set coverage problems as we verify experimentally in comparison to an optimal solver.
\begin{algorithm}
    \small
    \begin{algorithmic}[1]
    	\State{$i \gets 0$}
        \State{$\mathit{output} \gets []$}
        \While{$ i < K $}
            \State{$f_j^* \gets$ feature with maximum objective increase}
            \State{Remove tweets covered by $f_j^*$ from the data}
            \State{Add $f_j^*$ to $\mathit{output}$}
            \State{$i \gets i + 1$}
        \EndWhile
    \end{algorithmic}
    \caption{Greedy Optimization Framework}
    \label{alg:greedy}
\end{algorithm}

\end{enumerate}

\begin{table*}
\caption{Topics with five labeling hashtags and the resulting statistics of positively labeled tweets. The hashtags are selected by four independent human annotators, requiring an inner-annotator agreement of three annotators to permit a hashtag to be assigned to a topic set.}
\label{tab:toptag}
\small
\resizebox{\textwidth}{!}{%
\begin{tabular}{c|c|c|c|c|c|c|c}
\toprule  
& \textbf{Natural Disasters} &        \textbf{Social Issues} &          \textbf{Space} &             \textbf{Soccer} &    \textbf{Human Disasters} &          \textbf{Tennis} &             \textbf{Health} \\ 
\midrule
\textbf{\#Tweets} & 89,440 & 374,710 & 409,817 & 1,377,787 & 792,268 & 86,108 & 401,362 \\
\midrule
\multirow{5}{12mm}{\textbf{Sample} \textbf{Hashtags}} &
\#julio &           \#legalized &       \#houston &              \#fifa &   \#redefinenigeria &          \#nadal &  \#chanyeolvirusday \\
& \#tsunami2004 &        \#michaelbrown &        \#rocket &        \#halamadrid &         \#bombsquad &       \#usopen13 &         \#uniteblue \\
& \#chileearthquake &    \#berkeleyprotests &  \#meteorshower &  \#englandsoccercup &  \#malaysiaairlines &  \#wimbledon2o13 &       \#chikungunya \\ 
& \#hurricaneprep &  \#44millionabortions &     \#asteroids &           \#beckham &       \#notinmyname &    \#rafaelnadal &     \#ebolaresponse \\ 
     &  \#drought13 &         \#freetheweed &    \#astronauts &             \#messi &              \#mh17 &    \#murraynadal &     \#healthworkers\\ 
\bottomrule
\end{tabular}
}
\end{table*}

\section{Experimental Setup}

\subsection{Data Description.}
The data used in the following experiments is the crawled Twitter data retrieved using the Twitter streaming API for two years from 2013 to 2014.  Five general types of features present in each tweet are used as possible keywords for querying the search API: 
(1)~\textbf{From} -- the username of the tweet sender; 
(2)~\textbf{Location} -- provided by the users in their profiles;
(3)~\textbf{Hashtag} -- set of hashtags in each tweet starting with "\#" (if any);
(4)~\textbf{Mention} -- usernames of the users referenced in the tweet text with the ``@'' sign;
(5)~\textbf{Term} -- tokens in a tweet not matching (1)-(4).


Since the tweets are not labeled in their raw form, we need a method to properly label the data to be used with the proposed methods. Considering the large size of the data, manually labeling the tweets is practically impossible. Similar to the labeling methods in \cite{sanner:topical,Lin:2011:STA:2020408.2020476} the data is labeled for each of seven different topics using a set of user-curated hashtags (referred to as ``labeling hashtags'') specific to each topic. For each topic, the labeling hashtags are selected by four independent human annotators and each selected hashtag requires an inter-annotator agreement of at least three annotators to be included in the labeling hashtags used for labeling individual tweets. A subset of the hashtags used for labeling the tweets for each topic is shown in Table~\ref{tab:toptag} as well as the number of tweets labeled relevant for each topic. It should be noted that our proposed methods are independent of the labeling strategy and can be applied to any labeled dataset with different labeling methods.

During the preprocessing stage, duplicate tweets are removed from the dataset since they do not add any new information to the solution and are redundant. We focus on English tweets so that annotators can search and interpret the results of their chosen labeling hashtags.  To this end, we remove non-English tweets from the data during preprocessing. 
Since we use hashtags to label the tweets, tweets without any hashtags are also removed from the data since it's not possible to label them. Our dataset contains $135,910,871$ English tweets after the preprocessing stage. Due to the large size of the dataset, the size of the candidate keyword set for the queries is also very large ($17,996,431$ unique features in our training data). 
Thus, it is useful to apply frequency thresholds on the candidate query keywords to pare down the size of the candidate keyword set for efficient optimization purposes without losing critical features that can maximize topical coverage. We found that a modest frequency threshold of $100$ resulted in a reasonable feature set size of $73,887$ without removing important features (i.e., further decreasing the frequency threshold did not change the features selected for the API query).

We intentionally chose not to do stopword removal on the data. While the removal of common stopwords (e.g., it, the, a) is trivial, there can be more obscure and domain-related stopwords for each topic in our data due to the nature of the problem. As a general example, a term like ``rt'' can be considered a common stopword in Twitter data since it is present in each retweeted tweet. While it is possible to manually curate specific stopword collections for each topic, we expect our methods to be sophisticated enough to automatically detect and remove stopwords from their candidate solutions.

\subsection{Evaluation Framework.}
\begin{figure}
    \centering
    \includegraphics[width=\columnwidth]{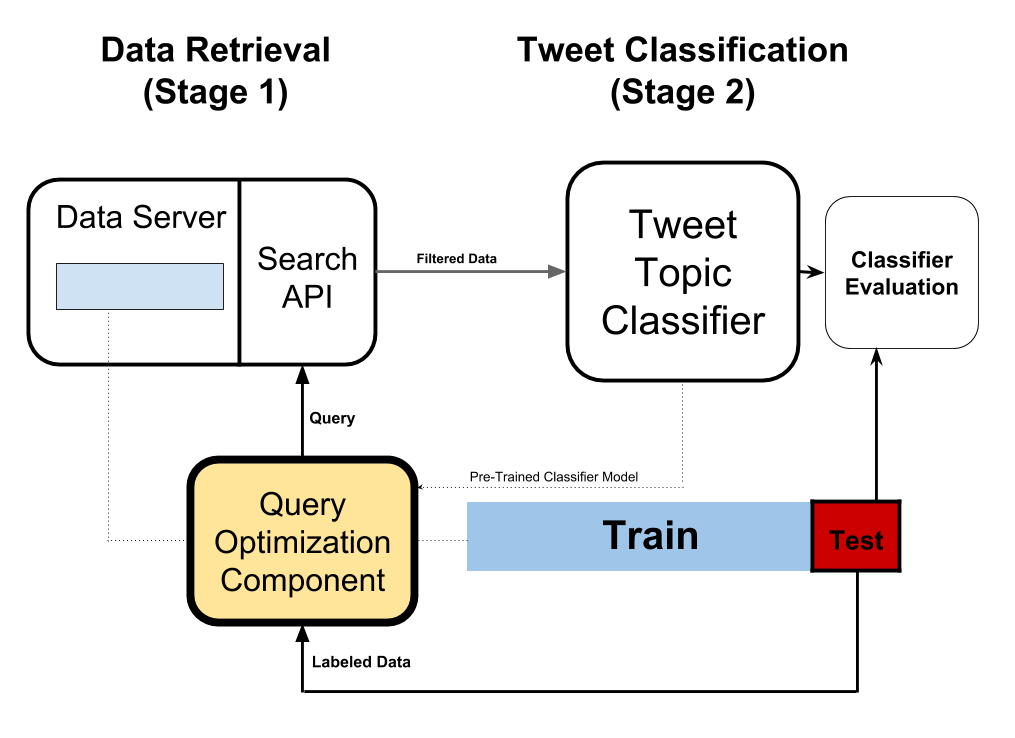}
    \caption{Overview of the evaluation framework. The data is randomly divided into five splits. The smaller split is used for deriving the query and evaluating the classifier and the rest of the data is used for training (to ensure the classifier is not trained on test data).}
    \label{fig:eval}
\end{figure}

Twitter recommends using ten keywords in each query.\footnote{\url{https://developer.twitter.com/en/docs/tweets/search/guides/standard-operators}} Given the fact that the maximum length of a query can be 500 characters and considering the fact that the API rejects overly complex queries, we set $K=20$ as the maximum number of features selected by our methods (up to $25$ characters on average per keyword).
We use 5-fold cross validation to evaluate the performance of our proposed methods. The tweets are randomly divided into five splits. At each iteration, we use one of the splits as the input for the query optimization component in Fig. \ref{fig:pipelines}. A query is derived using the data in this split which is used to filter the data in the rest of the data (consisting of the four remaining splits). The performance of the classifier is evaluated using the data from the initial smaller split. An overview of the evaluation process is depicted in Fig. \ref{fig:eval}.
We use a logistic regression classifier as the ranking model in the second stage. Similar to the labeling method, our choice of the classifier is independent of the proposed filtering methods.

Given the fact that we have different goals for the two stages of the framework, we evaluate different metrics at each stage to measure the performance of our proposed methods:

\begin{itemize}
    \item {\bf 1st Stage -- API query:} As previously mentioned in detail, the desired behavior of the first stage is maximizing the number of positive labeled tweets retrieved by the boolean query (high recall) while filtering the negative labeled tweets to have a more balanced distribution in the output (increased precision). Given these targets, we use recall and precision to measure the 1st stage performance.
    
    \item {\bf 2nd Stage -- Topical classification and ranking:} Since the output of the second stage classifier is presented to the user, it is important to have data with high precision specifically in the top ranked items in the output. This is why we evaluate the performance at this stage with more rank-centric metrics. The performance of the second stage is evaluated using Average Precision \cite{zhu:avep} and Precision@100 (users can reasonably examine the top-100 tweets given their short length).
\end{itemize}

In addition to the three proposed methods, we evaluated the performance on the firehose and also on an additional method that queries using the top-$K$ weighted features from the logistic regression classifier 
(referred to as the ``TopK'' method) as the baselines for evaluating the performance results of the ILP methods. 

\section{Results and Discussion}

\begin{table}
\centering
\caption{
Performance of query optimization methods for each topic.  Decimals are rounded to 3 digits and $95\%$ confidence intervals are shown.  Firehose does not use a restricted Query API and is simply provided for benchmark comparison; best results among TopK, CILP, WILP, and CAILP are bolded.}
\label{tab:res}
\npdecimalsign{.}
\nprounddigits{3}
\resizebox{\columnwidth}{!}{%
\begin{tabu}{l|[2pt]c|c|c|[2pt]c|c}
& \multicolumn{3}{c|[2pt]}{\textbf{First Stage Performance}} & \multicolumn{2}{c}{\textbf{Second Stage Performance}} \\
\toprule
Method &    \multicolumn{1}{c|}{Avg Retrieved} &   \multicolumn{1}{c|}{Recall} &  \multicolumn{1}{c|[2pt]}{Precision} & \multicolumn{1}{c|}{Test AveP} & \multicolumn{1}{c}{P@100}\\
\midrule
\multicolumn{6}{l}{\textbf{Natural Disaster}}\\
\midrule
Firehose &  96,079,424 &    1.000 &  0.001 $\pm$ 0.000 & 0.417 $\pm$ 0.004 & 0.772 $\pm$ 0.022 \\
TopK     &     235,231 &    0.340 $\pm$ 0.001 &  0.103 $\pm$ 0.000 & 0.373 $\pm$ 0.004 &  \textbf{0.928 $\pm$ 0.044} \\
CILP     &    9,580,565 &  \textbf{0.924 $\pm$ 0.001} &  0.007 $\pm$ 0.000 & 0.417 $\pm$ 0.004  & 0.750 $\pm$ 0.032\\
WILP     &  4,369,679  &  0.607 $\pm$ 0.030 &  0.010 $\pm$ 0.000 & 0.352 $\pm$ 0.101 & 0.754 $\pm$ 0.096 \\
CAILP   &   208,018 &  0.545 $\pm$ 0.023 &  \textbf{0.210 $\pm$ 0.093} &  \textbf{0.434 $\pm$ 0.009}  & 0.830 $\pm$ 0.057 \\
\midrule
\multicolumn{6}{l}{\textbf{Social Issues}}\\
\midrule
Firehose &  96,079,424 &            1.000 &  0.003 $\pm$ 0.000 & 0.678 $\pm$ 0.003 & 0.730 $\pm$ 0.035 \\
TopK     &     1,288,600 &  0.369 $\pm$ 0.001 &  0.086 $\pm$ 0.000 & 0.574 $\pm$ 0.005 & \textbf{0.741 $\pm$ 0.029} \\
CILP     &   9,718,098 &  0.732 $\pm$ 0.000 &  0.023 $\pm$ 0.000 & \textbf{0.684 $\pm$ 0.003} & 0.733 $\pm$ 0.035 \\
WILP     & 5,244,112 &  \textbf{0.839 $\pm$ 0.046} &  0.049 $\pm$ 0.008 & 0.603 $\pm$ 0.004  & 0.523 $\pm$ 0.018 \\
CAILP    &   353,827 &  0.808 $\pm$ 0.022 &  \textbf{0.720 $\pm$ 0.210} & 0.624 $\pm$ 0.005 &  0.714 $\pm$ 0.026 \\ 
\midrule
\multicolumn{6}{l}
{\textbf{Space}}\\
\midrule
Firehose & 96,079,424 &            1.000 &  0.003 $\pm$ 0.000 & 0.432 $\pm$ 0.002 & \textbf{0.746 $\pm$ 0.014} \\
TopK     & 592,934 &  0.203 $\pm$ 0.005 &  0.112 $\pm$ 0.003 & 0.533 $\pm$ 0.019 & 0.665 $\pm$ 0.016 \\
CILP     & 9,811,426 &  \textbf{0.900 $\pm$ 0.000} &  0.030 $\pm$ 0.000 & \textbf{0.871 $\pm$ 0.003} & 0.662 $\pm$ 0.008 \\
WILP     &  6,948,171 &  0.681 $\pm$ 0.030 &  0.032 $\pm$ 0.001 & 0.862 $\pm$ 0.005 & 0.652 $\pm$ 0.022 \\
CAILP    &    765,072 &  0.683 $\pm$ 0.053 &  \textbf{0.299 $\pm$ 0.059} & 0.794 $\pm$ 0.018 & 0.669 $\pm$ 0.013 \\
\midrule
\multicolumn{6}{l}{\textbf{Soccer}}\\
\midrule
Firehose &    96,079,424 &            1.000 &  0.011 $\pm$ 0.000 & 0.646 $\pm$ 0.001  & 0.700 $\pm$ 0.028 \\
TopK     &      2,759,962 &  0.244 $\pm$ 0.000 &  0.097 $\pm$ 0.000 &  0.399 $\pm$ 0.006 & 0.798 $\pm$ 0.037 \\
CILP     &      10,520,097 &  \textbf{0.901 $\pm$ 0.000} &  0.094 $\pm$ 0.000 & 0.648 $\pm$ 0.001 & 0.704 $\pm$ 0.029 \\
WILP     &    7,679,860 &  0.739 $\pm$ 0.007 &  0.106 $\pm$ 0.001 & \textbf{0.651 $\pm$ 0.001}  & 0.702 $\pm$ 0.036\\
CAILP    &   1,826,342 &  0.595 $\pm$ 0.289 &  \textbf{0.547 $\pm$ 0.252} & 0.599 $\pm$ 0.010 & \textbf{0.858 $\pm$ 0.016} \\
\midrule
\multicolumn{6}{l}{\textbf{Human Disasters}}\\
\midrule
Firehose &    96,079,424 &            1.000 &  0.007 $\pm$ 0.000 & 0.734 $\pm$ 0.002 & 0.602 $\pm$ 0.027\\
TopK     &    917,253 &  0.145 $\pm$ 0.003 &  0.100 $\pm$ 0.001 & 0.524 $\pm$ 0.014 & 0.672 $\pm$ 0.016\\
CILP     &     10,088,362 &  \textbf{0.868 $\pm$ 0.000} &  0.054 $\pm$ 0.000 & 0.739 $\pm$ 0.002 & 0.602 $\pm$ 0.033 \\
WILP     &   6,751,307 &  0.852 $\pm$ 0.027 &  0.080 $\pm$ 0.007 & \textbf{0.746 $\pm$ 0.003} & 0.614 $\pm$ 0.026 \\
CAILP    &845,854 &  0.678 $\pm$ 0.051 &  \textbf{0.531 $\pm$ 0.159} & 0.694 $\pm$ 0.006  & \textbf{0.684 $\pm$ 0.017}\\
\midrule
\multicolumn{6}{l}{\textbf{Tennis}}\\
\midrule
Firehose &     96,079,424 &            1.000 &  0.001 $\pm$ 0.000 & 0.851 $\pm$ 0.004 & 0.910 $\pm$ 0.015 \\
TopK      &   2,202,427 &  0.247 $\pm$ 0.001 &  0.008 $\pm$ 0.000 & 0.682 $\pm$ 0.021 & 0.922 $\pm$ 0.027 \\
CILP     &     9,518,243  &  \textbf{0.828 $\pm$ 0.000} &  0.006 $\pm$ 0.000 & \textbf{0.853 $\pm$ 0.004} & 0.910 $\pm$ 0.012 \\
WILP     &  5,047,539 &  0.665 $\pm$ 0.083 &  0.009 $\pm$ 0.001 & 0.840 $\pm$ 0.039  & 0.904 $\pm$ 0.007 \\
CAILP    &    69,615 &  0.513 $\pm$ 0.006 &  \textbf{0.512 $\pm$ 0.068} & 0.787 $\pm$ 0.026 & \textbf{0.940 $\pm$ 0.025} \\
\midrule
\multicolumn{6}{l}{\textbf{Health}}\\
\midrule
Firehose & 96,079,424 &            1.000 &  0.003 $\pm$ 0.000 & 0.532 $\pm$ 0.002 & 0.732 $\pm$ 0.033 \\
TopK     & 1,739,638&  0.324 $\pm$ 0.004 &  0.060 $\pm$ 0.000 & 0.439 $\pm$ 0.012 & 0.782 $\pm$ 0.016 \\
CILP     &  9,593,804 &  \textbf{0.818 $\pm$ 0.000} &  0.027 $\pm$ 0.000 & \textbf{0.537 $\pm$ 0.001} & 0.738 $\pm$ 0.032 \\
WILP     & 6,964,154 &  0.730 $\pm$ 0.039 &  0.034 $\pm$ 0.001 & 0.532 $\pm$ 0.002 & 0.738 $\pm$ 0.036 \\
CAILP    &528,421 &  0.621 $\pm$ 0.056 &  \textbf{0.446 $\pm$ 0.182} & 0.478 $\pm$ 0.009  & \textbf{0.792 $\pm$ 0.010} \\
\bottomrule
\end{tabu}%
}
\end{table} 

\begin{figure*}
\centering
\includegraphics[width=\textwidth]{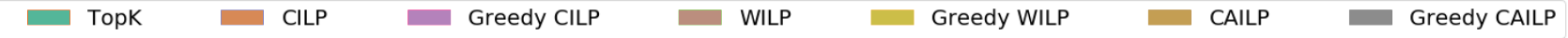}

\includegraphics[width=\textwidth]{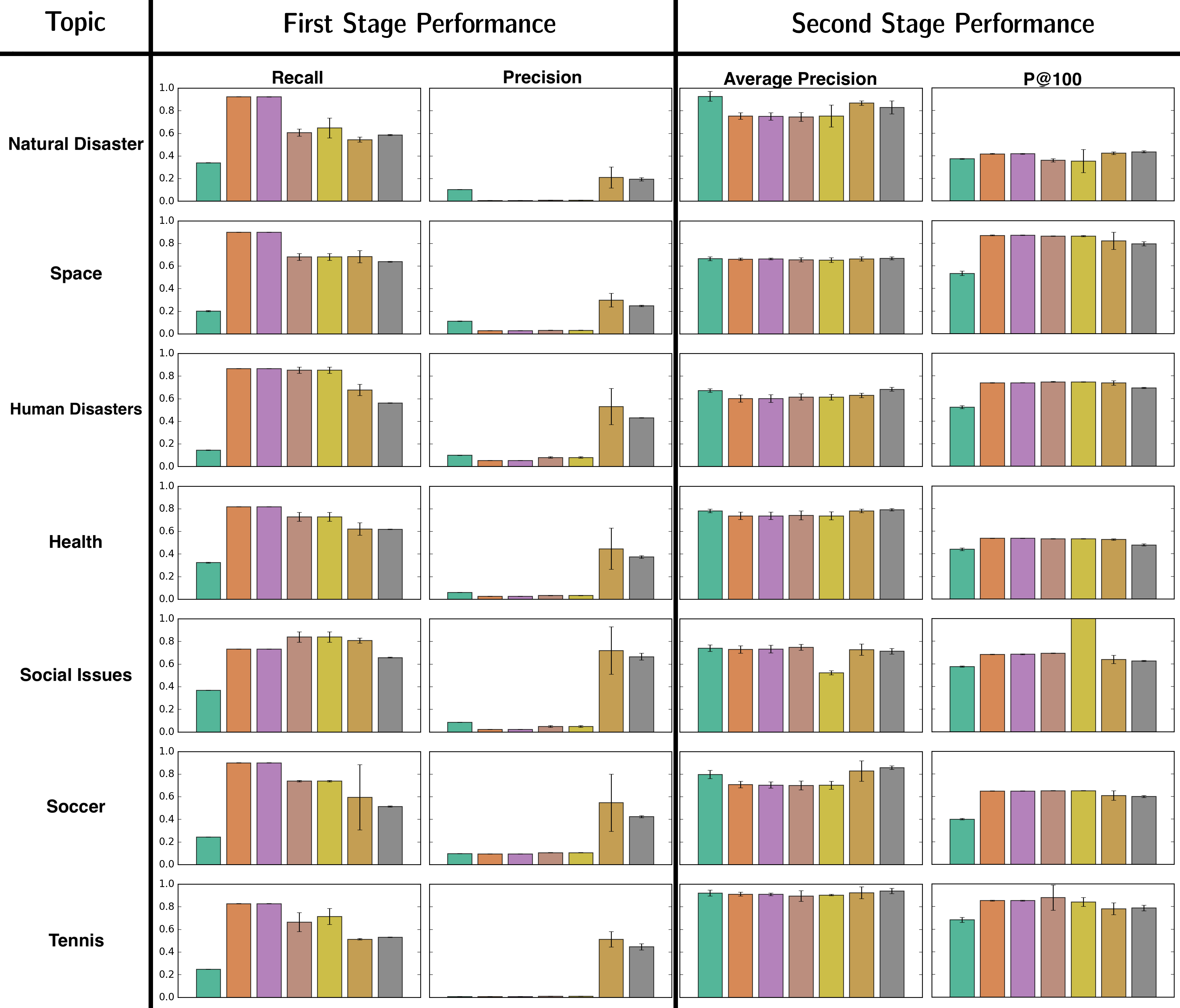}

\caption{
Performance comparison of different methods across all topics. As shown in Table~\ref{tab:res}, we remark that while CAILP typically retrieves less than 900,000 results of the roughly 96 million tweets of the Firehose, all other methods typically retrieve anywhere from 1-10 million results.  Though the TopK method performs well on metrics other than 1st stage recall, it typically retrieves more content than CAILP and performs notably worse than other methods on Precision@100.  While CAILP suffers on 1st stage recall (except compared to TopK), we remark that this is less important in the 2nd stage where only top-ranked content is presented to the user and evaluated.  CAILP achieves excellent precision at the 1st stage while achieving ranking performance in the 2nd stage that is either comparable or superior to all other baseline and ILP methods.  Finally, comparing (optimal) CAILP to Greedy CAILP, we note that there is little performance degradation due to greedy optimization; furthermore, the running time of greedy optimization shows that it is very efficient as shown in Figure~\ref{fig:time}.
}
\label{fig:all_res}
\end{figure*}

\subsection{Performance Comparison of Query Optimization Methods.}

Results for all proposed and baseline methods for the two stages are shown in Table~\ref{tab:res} and Figure~\ref{fig:all_res}. $95\%$ confidence intervals are shown for all results and calculated by applying each method on the five previously described splits of the tweet data. The ILP problems for the formulations are solved using the Gurobi solver\footnote{\url{http://www.gurobi.com}} which has shown the best performance among the available ILP solvers \cite{Meindl:2013:ACF:2612167.2612169}.  Figure~\ref{fig:all_res} shows a comparison of optimal vs. greedy ILP performance.

\vspace{1mm}
\noindent {\bf 1st Stage -- API query:} In this stage, we 
first examine Table~\ref{tab:res}.  We note that the Firehose achieves a recall of 1.0 by definition (it retrieves all data), but this comes at the expense of retrieving over 96 million tweets.  In contrast, the CAILP method typically has the fewest average retrieved tweets yet still performs competitively on recall and is undeniably the top performer for precision owing to its parsimonious retrieval.





The TopK method achieves higher recall than the Firehose but performs poorly on precision. This can be explained by the fact that the top-weighted features in a classifier are not learned with respect to coverage of the data. Therefore, while the features selected using this method are highly relevant to the topic, they can be redundant and will not guarantee full coverage of the topical content.

As for our proposed methods, using the information of the negative coverage of each feature in selecting the features does show significant improvements in the quality of the filtered data. Looking at the performance of the CAILP method, the recall may drop in comparison to the other two methods (0.64 mean recall vs 0.86 mean recall of the CILP method), however, there is a noticeable increase in the precision in the data retrieved by the CAILP method. The mean precision jumps from 0.048 in the WILP method to 0.48, which is significant. 

To summarize, the CILP and WILP methods recall the highest amount of positive labeled data, but they also fail to filter a considerable amount of irrelavnt data which results in noticeably low precision compared to the baseline TopK method. On the other hand, using the coverage of non-relevant tweets in the CAILP proved to be a good method for improving the queries; while the recall in the filtered data is not as high as the data from CILP and WILP, the higher precision gives us a more balanced distribution of the labeled data that can be queried in reasonable time within the API limits.

\vspace{1mm}
\noindent {\bf 2nd Stage -- Topical classification and ranking:} The second stage performance results in Table~\ref{tab:res} and Figure~\ref{fig:all_res} show that although the second stage of the ILP methods have significantly less data to work with than the Firehose, the second stage maintains (and for some topics improves) its ranking performance compared to the classifiers using the Firehose data. This can be explained by the fact that the majority of content filtered out by the API Queries in the first stage is actually non-topical.  Nonetheless, it is remarkable that CAILP performs comparable to the other baselines and ILP methods on the final ranking metrics given that it typically retrieves only a fraction of the data (rarely more than 900,000 tweets) that the other non-Firehose methods retrieve (up to 10,000,000 tweets).

\subsection{Performance of Greedy vs. Optimal Solvers.}

\begin{figure}
\centering
\includegraphics[width=.8\columnwidth]{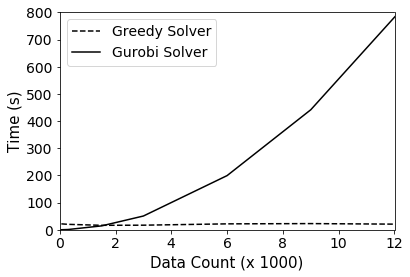}
\caption{Optimal vs. Greedy Gurobi ILP solve times.}
\label{fig:time}
\vspace{-3mm}
\end{figure}

In the previous sections we found that using solutions to the proposed ILP formulations output by the Gurobi solver can lead to good results on the filtered data. The problem with using a solver such as Gurobi is that each additional tweet adds a new constraint to the LP and the problem space will grow as the input data grows in size. Defining this constraint space and solving the ILP using Gurobi under these constraints can require infeasible processing times that render the methods unscalable.

We experimented with the greedy solver of Algorithm~\ref{alg:greedy} to see how it compared to the optimal Gurobi solver.  Specifically, we ran both solvers on the same data (for topic ``soccer'') with increasing data sizes and recorded the solve time; results are shown in Figure~\ref{fig:time}.  Here, the Gurobi solver performs better for small datasets, but as the size of the problem data increases, the time for Gurobi to solve the problem appears to grow exponentially. From our experiments, it took the Gurobi solver 82 minutes to terminate for an ILP with 300,000 tweets. In contrast, the greedy solver consistently terminates with a runtime measured in seconds.

While it is important to obtain a solution within a reasonable time, it is also important to ensure near-optimal results. To compare the results of the greedy solver with those of the optimal solver, we ran the greedy solver across all topics for each ILP method and recorded the performance metrics of the results. The results are plotted in Figure~\ref{fig:all_res}
with optimal and greedy bars side-by-side for easy comparison (see legend).

The performance results show that while the results of the greedy ILP solutions do not exactly match those of the the Gurobi solution, the greedy methods still manage to obtain similar performance to the optimal solution to the same ILP. 
As an example, while there is some performance degradation for CAILP in stage one when switching from the optimal to greedy solution, the difference is much less apparent in stage two.

\section{Related Work}
There has been a considerable amount of work to train complex topical classifiers and event detection systems for Twitter. However, the main focus in most of these works is on a pre-cached set of data previously acquired by querying the API under more simple rules and queries; this results in the retrieval of excessive non-relevant content that can be avoided through the application of a more sophisticated data collection method. 
For example, a number of works 
\cite{Lin:2011:STA:2020408.2020476}, \cite{Yang:2014:LHT:2623330.2623336} and 
\cite{Magdy_adaptivemethod} propose methods for following topics on Twitter, however, all of them query the API for raw and unfiltered data which may waste bandwidth and resources to retrieve large volumes of unrelated content that we tried to avoid in this work. 

As for existing work on searching microblog services such as Twitter, Hao et al.~\cite{hao:queryless} introduce methods using query expansion to retrieve related tweets for a user's interest based on a tweet selected by the user as a seed and generating new queries based on this tweet. Similarly, Li et al.\cite{tedas} have implemented a crawler in their work which uses query expansion methods to iteratively retrieve tweets relevant to a specific topic (crime and disaster event tweets). New queries are generated based on the results of previous queries until termination.  In fact, there are a multitude of additional works leveraging dynamically evolving query approaches~\cite{becker2012identifying,Wang:2013:EHA:2492517.2492624,Li:2014:RHR:2600428.2609618,Zheng:2017:SET:3132847.3132968}.  We simply remark that our objective in this work is complementary --- we aimed simply to pre-optimize a single ``best'' query given labeled data; one could view our method as an initial query to be expanded upon by these query reformulation methods if search API bandwidth permitted it.

The closest work that uses maximum coverage in a similar context is by Saha and Getoor~\cite{saha} who leverage it to find blogs that have maximal relevancy to a list of topics.
While they focus on efficient, bounded approximations of maximum coverage similar to CILP in an online streaming environment, 
we focus our contributions on improved variants (CAILP and WILP) of the underlying maximum coverage ILP in an offline optimization framework.  It would be interesting future work to combine these two research threads.

\section{Conclusion}


In this work, we introduced maximum coverage optimization strategies to select search API queries that achieve high precision and recall for topical data at a first stage of retrieval to be used for ranked retrieval at a second stage.
The proposed CAILP method provided strong performance on all metrics at both the first and second stage of retrieval (including vs. TopK and the Firehose) while requiring only a fraction of the communication bandwidth of other methods.  Furthermore, fast greedy optimization of CAILP yielded little performance degradation compared to the optimal solver.


In concluding, we remark that there are a variety of potential applications of this work beyond Twitter topic classifiers.  Overall, as the demand for free, real-time data faces the reality of limited search APIs, the need for ways to retrieve content relevant to application needs will only grow.  Leveraging methods such as those proposed in this paper not only help minimize data costs to end users, but also limit the communication bandwidth required.  Future work may seek to improve the existing objectives and optimization methods, examine other platforms beyond Twitter, and investigate other API-restricted machine learning problems (e.g., regression, time series forecasting) as well as solutions that are tailored to specific machine learning paradigms (e.g., random forest or deep neural networks).

\bibliographystyle{SIAM}
\bibliography{bibliography}

\end{document}